\documentclass[12pt]{iopart}

\usepackage{amsfonts}
\usepackage{amssymb}
\usepackage{iopams}
\usepackage{graphicx}
\usepackage{color}
\usepackage{latexsym}
\usepackage{array}
\usepackage{soul}

\begin{document}

\title[Active particles with fractional rotational Brownian motion]{Active particles with fractional rotational Brownian motion}

\author{Juan Ruben Gomez-Solano$^{1,*}$ and Francisco J. Sevilla$^{1}$}

\address{$^1$Instituto de F\'isica, Universidad Nacional Aut\'onoma de M\'exico,\\
Apdo. Postal 20-364, 01000, Ciudad de M\'exico, M\'exico}

\ead{$^{*}$r\_gomez@fisica.unam.mx}

\vspace{10pt}
\begin{indented}
\item[]January 2020
\end{indented}

\begin{abstract}
We study the two-dimensional overdamped motion of an active particle whose orientational dynamics is subject to fractional Brownian noise, whereas its position is affected by self-propulsion and Brownian fluctuations. From a Langevin-like model of active motion with constant swimming speed, we derive the corresponding Fokker-Planck equation, from which we find the angular probability density of the particle orientation for arbitrary values of the Hurst exponent that characterizes the fractional rotational noise. We provide analytical expressions for the velocity autocorrelation function and the translational mean-squared displacement, which show that active diffusion effectively emerges in the long-time limit for all values of the Hurst exponent. The corresponding expressions for the active diffusion coefficient and the effective rotational diffusion time are also derived. Our results are compared with numerical simulations of active particles with rotational motion driven by fractional Brownian noise, with which we find an excellent agreement.
\end{abstract}

%
%
%
%
%


\section{Introduction}\label{intro}

Active matter is composed of non-equilibrium entities capable of converting autonomously  the energy of their surroundings into directed motion or into mechanical work~\cite{bechinger2016}. This definition encompasses a broad variety of systems such as locomotive animals, motile microorganisms, microtubules and actin filaments in the cytoskeleton of eukaryotic cells, biomolecular motors, active colloids and nanomotors~\cite{ramaswamy2010}. In particular, one type of active systems that has been the subject of intensive research in recent years are the so-called microswimmers. These are micron- or submicron-sized particles that propel themselves through fluid environments at vanishing Reynolds numbers ~\cite{elgeti2015}. Most microswimmers exhibit patterns of locomotion characterized by a persistent ballistic motion at sufficiently short time scales, whereas its swimming direction is not fixed but strongly affected by random changes in the particle orientation~\cite{bechinger2016,taktikos2013}. Such orientational changes can be caused by, e.g., stochastic polymorphic transformations during the run-and-tumble dynamics of flagellated bacteria~\cite{darnton2007}, or due to thermal fluctuations from the surrounding fluid, as those observed for self-propelled colloids~\cite{howse2007}. For time scales much larger than the typical reorientation time (\emph{persistence time}), the randomization of the microswimmer orientation often leads to an effective diffusive behavior with a diffusion coefficient higher than expected under thermal equilibrium conditions \cite{howse2007,saragosti2012,cates2013}.

One of the most-widely studied models of active motion of microswimmers is the so-called active Brownian particle (ABP). It consists of a disc or a sphere moving with a prescribed swimming velocity, where the direction of self-propulsion is determined by a well-defined orientation vector that changes with time by rotational diffusion~\cite{tenhagen2011,pototsky2012,redner2013,bialke2013,sevilla2014,sevilla2015,basu2018}. Both the translational and the angular motion of the ABP are overdamped and the inverse of the rotational diffusion coefficient sets the reorientation time above which the fluctuations of particle orientation become decorrelated. Despite its simplicity, the ABP model captures the main dynamical features of many active colloidal systems \cite{kurzthaler2018}. For instance, it exhibits a swimming-persistence length, determined by the product of the propulsion speed and the rotational diffusion time, whereas a translational diffusion coefficient that depends quadratically on the swimming velocity emerges in the long-time limit. Such a behavior is consistent with an exponential decay of the swimming-velocity autocorrelation function, as experimentally measured in several active colloidal suspensions~\cite{howse2007,bregulla2015,gomezsolano2017}.  
Furthermore, the ABP model is also the basis for the theoretical investigation of intricate phenomena in active matter, such as self-propelled motion in presence of external potentials~\cite{solon2015,vachier2019,woillez2019}, external flows~\cite{tenhagenpre2011,zoettl2012,li2017}, under geometrical confinement~\cite{hu2017,duzgun2018,wagner2019,caprini2019}, as well as collective phenomena of self-propelled particles, e.g. dynamical clustering and motility-induced phase separation \cite{cates2013,redner2013,bialke2013,wysocki2014,stenhammar2014,richard2016}. Moreover, the ABP model has allowed the analysis of stochastic thermodynamics and symmetry properties of extreme fluctuations in active systems~\cite{speck2016,pietzonka2016,falasco2016,gaspard2017,shankar2018}. 

In this paper, we investigate the active motion of self-propelled particles whose orientation is driven by long-ranged correlated noise. While most of the Langevin models of active particles consider the effect of thermal fluctuations by means of delta-correlated Gaussian noises, only a few works have addressed the role of long-lived correlations in the rotational motion~\cite{hu2017,peruani2007,gosh2015,debnath2016,narinder2018,sevilla2019}. In fact, memory effects caused by colored noise are a common feature of the orientational motion of active particles in complex media~\cite{narinder2018,gomezsolano2016,lozano2018,lozano2019,saad2019,narinder2019}. In particular, our work is motivated by the occurrence of anomalous diffusion in several examples of soft matter systems, where spatial heterogeneities lead to a non-linear monotonic growth with time of the mean-square displacement \cite{chepizhko2013}. 
For instance, anomalous diffusion arises in the translational motion of granules within the cytoplasm~\cite{tolic2004}, colloidal beads in entangled actin filament networks~\cite{wong2004} or wormlike micelles~\cite{jeon2013}, and vacuoles in highly motile amoeboid cells~\cite{thapa2019}.
Moreover, there is direct evidence of anomalous \emph{rotational} diffusion in the orientational dynamics of single molecules of polymer melts near the glass transition~\cite{deschenes2001},  the rotational subdiffusion of proteins~\cite{cote2010} and the angular fluctuations of spherical colloidal probes suspended in polymer solutions \cite{andabloreyes2005,gutierrezsosa2018}.
Among the distinct approaches to describe anomalous diffusion~\cite{oliveira2019}, here we focus on fractional Brownian noise~\cite{qian2003} in order to capture the effect of long-ranged temporal correlations in the directional changes of a self-propelled particle.

The paper is organized as follows.
In Section \ref{Sect:Model} we present a model that describes the two-dimensional (2D) motion of a self-propelled particle subject to fractional rotational Brownian noise. The Fokker-Planck equation for the one-particle probability density, and the explicit solution for the angular density are shown in Section~\ref{Sect:FP}. Then, in Section~\ref{Sect:AD} we derive the analytical expressions for the swimming-velocity autocorrelation function and the corresponding mean-squared displacement. In the same Section we analyze the  different time scales that arise due to the persistent or antipersistent nature of the rotational noise, as well as the emergence of an active diffusion coefficient in the asymptotic limit. Finally, in Section~\ref{Sect:Conc} we summarize the main results of our work and make some further remarks.

\section{Model of active motion with fractional rotational Brownian noise}\label{Sect:Model}

\subsection{\label{subSect:LangevinEquation}Langevin equation}

We consider a particle that self-propels and rotates in a two-dimensional domain with swimming velocity $\boldsymbol{v}_{s}(t) = v_{s}(t)\hat{\boldsymbol{v}}(t)$, where $\hat{\boldsymbol{v}}(t)=\left[ \cos \varphi (t),\sin \varphi
(t)\right]$ is a unit vector defined by the angle $\varphi (t)$ between the direction of swimming and the horizontal axis of a given Cartesian reference frame, whereas $v_{s}(t)$ is the swimming speed. Generally, an active particle is subject to the fluctuations of a possibly complex surrounding fluid, and also to strongly correlated active fluctuations that originate from the particle's internal mechanisms that result in self-propulsion (see for instance Ref. \cite{figueroa2018}). In many cases of interest, these two kinds of fluctuations are strongly coupled by hydrodynamic effects, by the particular mechanism of self-propulsion (as in Janus particles), or by both of them. In this paper, we address the case for which thermal fluctuations are weakly coupled to active ones, and their statistical properties are assumed to be independent of each other. The thermal velocity fluctuations $\boldsymbol{\xi }_{T}(t)$ affect the translational part of motion and are characterized by a fixed temperature $T$.
These are modeled as white noise, i.e., $\langle \boldsymbol{\xi }_{T}(t)\rangle =\boldsymbol{0}$ and $\langle 
\xi _{i,T}(t)\xi _{j,T}(s)\rangle =2D_T\delta (t-s)\delta_{i,j}$, where the subscripts $i,j$ denote the Cartesian components $x$, $y$, of a two-dimensional vector and $D_T$ the diffusion constant given by $k_{B}T \mu$, with $\mu$ the particle mobility. On the other hand, strongly correlated active fluctuations, $\xi _{R}(t)$, which originate from the internal mechanism of self-propulsion that gives rise to persistent motion, drive the angular velocity of the particle orientation. Such active fluctuations are modeled as a colored Gaussian noise with vanishing mean, $\langle \xi_{R}(t)\rangle =0$, and autocorrelation function
$\langle \xi _{R}(t)\xi _{R}(s)\rangle =\omega(t,s)$. In this way, the overdamped dynamics of this active particle is described by the following Langevin equations 
\numparts
\begin{eqnarray}\label{modelo}
\frac{d{}}{dt}\boldsymbol{x}(t)& = &v_{s}(t)\,\hat{\boldsymbol{v}}(t)+\boldsymbol{
\xi}_{T}(t),  \label{LangevinPosition} \\
\frac{d}{dt}\varphi(t)&= &\xi_{R}(t),  \label{LangevinDirection}
\end{eqnarray}
\endnumparts
where $\boldsymbol{x}(t)$ denotes the particle position in two dimensions at time $t$, given in Cartesian coordinates by the vector $\boldsymbol{x}(t)=[x(t),y(t)]$. 
In the following, we restrict our analysis to the case for which the swimming speed remains constant over time, $v_s(t) = v_0$ , and where active angular fluctuations are a stochastic stationary process, i.e. they satisfy $\omega(t,s)=\omega(t-s)$. In particular, we investigate the situation for which $\xi_{R}(t)$ corresponds to a continuous-time \emph{fractional Gaussian noise} \cite{qian2003}, which endows the rotational motion with persistent or antipersistent correlations. Specifically, the autocorrelation function of $\xi_{R}(t)$ is 
given by
\begin{equation}\label{fGnoise}
 \omega(t-s) =2HD_{H}\vert t-s\vert^{2H-1}\left[\frac{2H-1}{\vert t-s\vert}+2 \delta(t-s)\right].
\end{equation}
In Eq.~(\ref{fGnoise}), $0 < H < 1$ stands for the so-called Hurst exponent that characterizes the long-ranged time correlation of the active fluctuations, while $D_{H}$ (with units of [time]$^{-2H}$) depicts their amplitude. Such a long-ranged autocorrelation function models long-term memory effects of a variety of stochastic phenomena that have stationary increments and exhibit self-similarity, i.e., processes for which their probability density distribution is invariant under changes of scale \cite{kou2004}. These processes generalize the standard Brownian motion by means of a power-law decay that takes into account the effect of previous values of the noise at time $s \le t$ on the current one at $t$.
For $0 < H < \frac{1}{2}$, the stochastic process $\xi _{R}(t)$ is negatively correlated (anticorrelated\footnote{Broadly speaking, this means that the particle changes its direction of motion from clockwise to counter-clockwise in an stochastic manner.}) and describes an \emph{antipersistent} orientational dynamics, whereas for $\frac{1}{2} < H < 1$, the motion is positively correlated (\emph{persistent}), i.e. if $t \neq s$, then $\omega(t-s) < 0$ or $\omega(t-s) > 0$, respectively. These two distinct behaviors have deep consequences in the pattern of motion of the active particle, as it will be unveiled in the following sections. The specific value $H = \frac{1}{2}$ corresponds to the case of Gaussian white noise, $\omega(t-s) = 2 D_{1/2} \delta(t-s)$.
Furthermore, the process $\varphi(t) = \int_0^t \xi_{R}(t') dt'$ is a continuous-time \emph{fractional Brownian motion} with zero mean, i.e. $\langle \varphi(t) \rangle = 0$, and autocorrelation function
\begin{equation}\label{fGmotion}
 \langle \varphi(t) \varphi(s) \rangle =D_{H}\left( |t|^{2H} + |s|^{2H} - |t-s|^{2H} \right),
\end{equation}
where the brakets denote an ensemble average over different realization of the fractional Brownian noise $\xi_{R}(t)$. Eq. (\ref{fGmotion}) yields the following expression for the mean-squared angular displacement
\begin{equation}\label{msdfGmotion}
 \langle \varphi(t)^2 \rangle = 2D_H t^{2H},
\end{equation}
where $t \ge 0$ and $\varphi$ is defined in the unrestricted domain $(-\infty,\infty)$. Eq.~(\ref{msdfGmotion}) reduces to the case of rotational Brownian diffusion for $H = \frac{1}{2}$, whereas the values $0 < H < \frac{1}{2}$ and $\frac{1}{2}< H < 1$ correspond to rotational subdiffusion and superdiffusion, respectively. 

\begin{figure*}
\includegraphics[width=\textwidth]{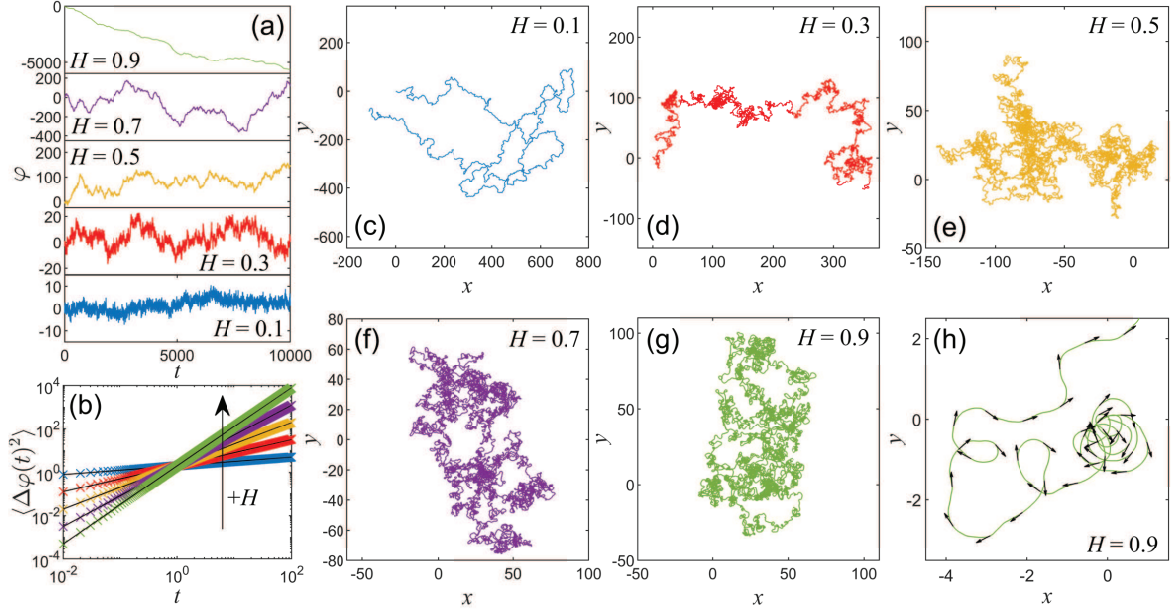}
 \caption{(a) Examples of the stochastic time evolution of the angle $\varphi(t)$ obtained from numerical simulations of fractional Brownian motion for different values of the Hurst exponent $H$. From bottom to top: $H = 0.1, 0.3, 0.5, 0.7, 0.9$. (b) Corresponding mean-squared angular displacements $\left\langle \Delta\varphi(t)^{2}\right\rangle$. The symbols correspond to the numerical results obtained from the simulated trajectories in (a) whereas the solid lines are computed from Equation~(\ref{msdfGmotion}). The resulting 2D trajectories of the particle position $(x,y)$ are plotted for (c) $H = 0.1$, (d) $H=0.3$, (e) $H = 0.5$, (f) $H = 0.7$, and (g) $H = 0.9$. All trajectories start at $[x(0)=0, y(0) = 0]$. (h) Expanded view of the active trajectory in (b), showing the loops resulting from the superdiffusive angular motion for $H = 0.9$. The arrows represent the instantaneous orientation $\hat{\boldsymbol{v}}(t) = [\cos \varphi(t), \sin \varphi(t)]$ at different times $t$.}
 \label{fig:trajectories}
\end{figure*}

\subsection{Numerical analysis}

In order to gain insight into the statistics of active particles subject to fractional rotational Brownian motion, we have simulated stochastic trajectories evolving according to Eqs.~(\ref{LangevinPosition})-(\ref{LangevinDirection}) for different values of $H$. To this end, fractional Brownian motion with autocorrelation function given by (\ref{fGmotion}), is generated using the circulant embedding method of the covariance matrix \cite{dietrich1997}, whereas the 2D particle position is solved by means of an Euler-Cromer scheme with time step $10^{-3}D_H^{-\frac{1}{2H}}$.
In the numerical results presented throughout the paper, velocities, times cales, length scales, and translational diffusion coefficients are normalized by $v_0$, $\tau_{H}\equiv D_H^{-\frac{1}{2H}}$, $\ell_{H}\equiv v_0 D_H^{-\frac{1}{2H}}$, and $\mathcal{D}_{H}\equiv v_0^2 D_H^{-\frac{1}{2H}}$, respectively. In particular, $\tau_{1/2} \equiv D_{1/2}^{-1}$ and $\ell_{1/2} \equiv v_0 D_{1/2}^{-1}$ correspond to the rotational diffusion time and the swimming-persistence length for active Brownian motion driven by rotational diffusion ($H = \frac{1}{2}$), respectively.

In Fig.~\ref{fig:trajectories}(a), we show some examples of the temporal evolution of the angle $\varphi(t)$, in the extended domain  $(-\infty,\infty)$, for different values of the Hurst exponent over the time interval $0 \le t \le 10^{4}\, D_H^{-\frac{1}{2H}}$. The corresponding mean-squared angular displacements are plotted in Fig.~\ref{fig:trajectories}(b), thus showing agreement with the expression given in Eq.~(\ref{msdfGmotion}). 2D trajectories of the active particle position $[x(t),y(t)]$, resulting from the fractional rotational Brownian motion, are shown in Figs.~\ref{fig:trajectories}(c)-(h).
For the sake of simplicity and in order to better appreciate the separate effect of fractional rotational Brownian noise on the 2D active motion, here we focus on the case without translational fluctuations, i.e., $D_{T}=0$. We want to clarify this is not an approximation, but rather a consequence of the separate dynamics given by Eqs.~(\ref{LangevinPosition})-(\ref{LangevinDirection}), as is shown in the next Section \ref{Sect:FP}. For $0<H<\frac{1}{2}$, the anti-persistence of the stochastic rotational dynamics leads to a highly persistent translational motion, as seen in Figs.~\ref{fig:trajectories}(c) and \ref{fig:trajectories}(d) for $H=0.1$ and $0.3$ respectively. This translates into actual persistence lengths much larger than the one expected for $\delta$-correlated rotational diffusion, $\ell_{1/2}$. This effect vanishes for $H=\frac{1}{2}$, for which the rotational dynamics results in diffusive translational motion with an effective active diffusion coefficient $D_{1/2}^{\mathrm{eff}} = \frac{1}{2} \mathcal{D}_{1/2} = \frac{1}{2}v_0^2 D_{1/2}^{-1}$. Then, for observation times much larger than $\tau_{1/2}=D_{1/2}^{-1}$, the swimming persistence is lost and the particle performs an effective memoryless random walk, as is seen in Fig.~\ref{fig:trajectories}(e). Similarly, for $\frac{1}{2}<H<1$, an \emph{active random walk} also emerges at time scales much larger than $D_H^{-\frac{1}{2H}}$, as shown in Figs~\ref{fig:trajectories}(f) and~\ref{fig:trajectories}(g) for $H = 0.7$ and $H=0.9$, respectively. However, a close inspection of the active trajectories reveals that the short-time motion is qualitatively distinct from the active Brownian motion for $H = 0.5$: looped trajectories are formed as $H$ increases and become more conspicuous as $H$ approaches the value 1, see Fig.~\ref{fig:trajectories}(g) for $H = 0.9$. We point out the different nature of such looped trajectories from the ones developed by chiral self-propelled particles driven by a constant torque~\cite{loewen2016} or by biased orientational fluctuations, for which the sense of rotation remains fixed over time. Instead, the trajectories obtained in this paper resemble the stochastic circular orbits that emerge in active colloids moving in viscoelastic fluids~\cite{narinder2018} and the meandering and chaotic motion predicted for self-phoretic particles at large P\'eclet numbers~\cite{hu2019}.

\section{Statistics of active motion}\label{Sect:FP}

\begin{figure*}
\includegraphics[width=\textwidth]{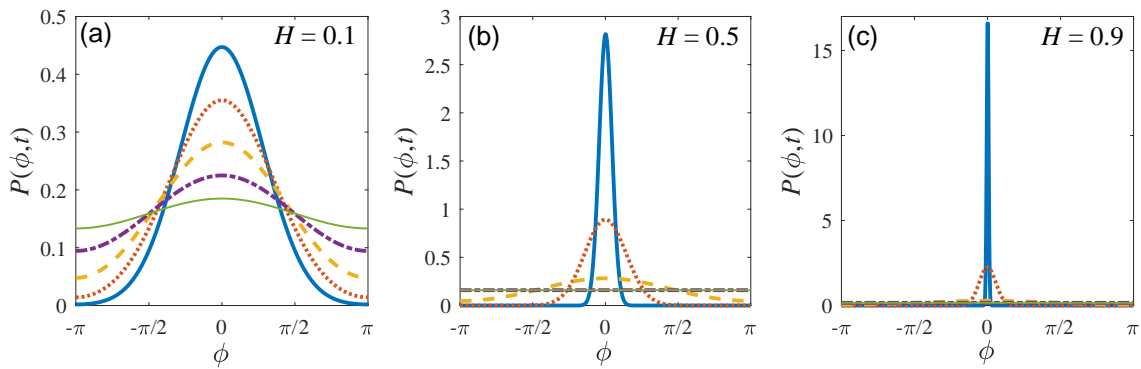}
 \caption{Profiles of the probability density function of the particle orientation angle $\phi$, as given in Eq. (\ref{eq:angdistr}) for the initial angular distribution $P(\phi',0) = \delta(\phi')$, for different values of $H$: (a) $H = 0.1$ (subdiffusive regime), (b) $H = 0.5$ (diffusive regime), (c) $H = 0.9$ (superdiffusive regime), at different times. From sharper to flatter curves: $t = 0.01 D_H^{-\frac{1}{2H}}$ (thick solid line), $t = 0.1 D_H^{-\frac{1}{2H}}$ (dotted line), $t = D_H^{-\frac{1}{2H}}$ (dashed line), $t = 10 D_H^{-\frac{1}{2H}}$ (dotted-dashed line) and $t = 100 D_H^{-\frac{1}{2H}}$ (thin solid line).}\label{fig:pdfangle}
\end{figure*}

\subsection{Fokker-Planck equation}

The Fokker-Planck equation for the one-particle probability density $p({\boldsymbol{x}},\varphi ;t)\equiv \langle \delta [{\boldsymbol{x}%
}-{\boldsymbol{x}}(t)]\delta [\varphi -\varphi (t)]\rangle$ that corresponds to the Langevin equations (\ref{modelo}), can be derived by standard methods, see for instance 
\cite{sevilla2014,sevilla2015}. By taking the derivative with respect to time $t$, we get 
\begin{eqnarray}\label{eq:derivFP}
\frac{\partial}{\partial t} p({\boldsymbol{x}},\varphi;t) & = & - v_0 \hat{\boldsymbol{v}} \cdot \nabla p({\boldsymbol{x}},\varphi;t) \nonumber\\
 & & - \nabla \cdot \langle \boldsymbol{
\xi}_{T}(t) \delta[\boldsymbol{x} - \boldsymbol{x}(t)] \delta[\varphi - \varphi(t)]\rangle \nonumber\\ 
 & & - \frac{\partial}{\partial \varphi} \langle \xi_R(t) \delta[\boldsymbol{x} - \boldsymbol{x}(t)] \delta[\varphi - \varphi(t)]\rangle,
\end{eqnarray}
where $\nabla = \left(\frac{\partial}{\partial x}, \frac{\partial}{\partial y} \right)$ and the brackets stand for an average over realizations of both the translational and rotational noises,  $\boldsymbol{\xi}_{T}(t)$ and $\xi_R(t)$, respectively. The second and the third terms on the right-hand side of Eq.~(\ref{eq:derivFP}), namely, the mean value of the product of the functional $\delta[\boldsymbol{x} - \boldsymbol{x}(t)] \delta[\varphi - \varphi(t)]$ with the Gaussian noises $\boldsymbol{
\xi}_{T}(t)$ and $\xi_R(t)$, respectively, can be evaluated by applying the Furutsu-Novikov theorem~\cite{furutsu1963,novikov1965}. A straightforward calculation leads to the Fokker-Planck equation for the Langevin model (\ref{modelo})
\begin{equation}\label{eq:FPfGn}
\frac{\partial }{\partial t}p({\boldsymbol{x}},\varphi ;t)+v_{0}\hat{
\boldsymbol{v}} \cdot \nabla p({\boldsymbol{x}},\varphi;t)= \\ D_{T}\nabla^{2}p({\boldsymbol{x}},\varphi ;t) 
+\Omega(t)\frac{\partial^{2}}{\partial\varphi^{2}}p({\boldsymbol{x}},\varphi;t), 
\end{equation}
where $\Omega(t)=\int_{0}^{t}ds\, \omega(s)$ for an arbitrary stationary correlation function $\omega(t)$ of the rotational noise $\xi_R(t)$. We point out that although the process $\varphi(t)$ is not strictly Markovian, its Gaussian nature allows the representation of the associated Fokker-Planck equation for the probability density $p({\boldsymbol{x}},\varphi ;t)$ as a time-dependent Markov process~\cite{haenggi1982}.

In the case of fractional Gaussian noise, the correlation function given in Eq.~(\ref{fGnoise}) leads to
\begin{equation}\label{eq:intcorrfGnoise}
\Omega(t)=2HD_{H}t^{2H-1}.
\end{equation}
Note that the parameter $\Omega(t)$ given by Eq.~(\ref{eq:intcorrfGnoise}) plays the role of a time-dependent rotational diffusion coefficient in the Fokker-Planck equation (\ref{eq:FPfGn}), where a Brownian-like rotational diffusion coefficient, $\Omega(t) = D_{1/2}$, is recovered for $H = \frac{1}{2}$.

We point out here that the net diffusion of a free active particle described by Eq. (\ref{eq:FPfGn}) can be split into the actively-induced free diffusion due to orientational fluctuations, and the free diffusion induced by thermal fluctuations. By writing
\begin{equation}
    p(\boldsymbol{x},\varphi,t)=\int d^{2}x^{\prime}G(\boldsymbol{x}-\boldsymbol{x}^{\prime};t)p_{a}(\boldsymbol{x}^{\prime},\varphi,t),
\end{equation}
where $G(\boldsymbol{x};t)$ is the bivariate Gaussian distribution that solves the diffusion equation $\partial_{t}G(\boldsymbol{x};t)=D_{T}\nabla^{2}G(\boldsymbol{x};t),$
it can be shown that the diffusion induced by the orientational fluctuations is described by
\begin{equation}
    \frac{\partial }{\partial t}p_{a}({\boldsymbol{x}},\varphi ;t)+v_{0}\hat{
\boldsymbol{v}} \cdot \nabla p_{a}({\boldsymbol{x}},\varphi;t)=\Omega(t)\frac{\partial^{2}}{\partial\varphi^{2}}p_{a}({\boldsymbol{x}},\varphi;t), \label{eq:ActiveFPfGn}
\end{equation}
where $p_{a}(\boldsymbol{x},\varphi;t)$ gives the probability density of finding a particle at $\boldsymbol{x}$, moving in the direction $\varphi$ at time $t$, due to self-propulsion only. Notice that~(\ref{eq:ActiveFPfGn}) can be obtained from Eq.~(\ref{eq:FPfGn}) by simply putting $D_{T}=0$.

\subsection{Angular probability density function}\label{subsect:angPDF}

By integrating Eq.~(\ref{eq:FPfGn}) with respect to $\boldsymbol{x}$ over the entire two-dimensional spatial domain, we find the Fokker-Planck equation for the probability density of the angle $\varphi$ at time $t$, i.e., for $P(\varphi,t) = \int_{-\infty}^{\infty} \int_{-\infty}^{\infty} dx\,dy\, p_{a}(\boldsymbol{x},\varphi;t)$ 
\begin{equation}\label{eq:FPangle}
    \frac{\partial}{\partial t}P(\varphi,t) = \Omega(t)\frac{\partial^2}{\partial \varphi^2}P(\varphi,t).
\end{equation}
This corresponds to the diffusion equation with time-dependent rotational diffusion coefficient $\Omega(t)$. The periodicity of $P(\varphi,t)$ with respect to the variable $\varphi$ is imposed by requiring that $P(\varphi,t)=P(\varphi+2\pi,t)$. We restrict the domain of description for the particle orientation by introducing the new angle $\phi$, which is defined in the interval $(-\pi,\pi]$.
Then, the probability density that a single active particle transits from moving along the direction $\phi^{\prime}$ at time $t^{\prime}$ to move along the direction $\varphi$ in the time interval $t-t^{\prime}$, $\mathcal{P}(\phi,t-t^{\prime}\vert\phi^{\prime})$, is given by the solution of Eq.~(\ref{eq:FPangle}) for time $t \ge t^{\prime}$. Given the initial condition $\mathcal{P}(\phi, 0\vert\phi^{\prime}) = \delta(\phi-\phi^{\prime})$, such a solution is found by writing $P(\phi,t)$ as a Fourier series $\sum_{n}c_{n}(t-t^{\prime},\phi^{\prime})e^{in\phi}$, where the coefficients $c(t-t^{\prime},\phi^{\prime})$ of such expansion are explicitly determined from Eq. (\ref{eq:FPangle}) and the initial condition, namely
\begin{eqnarray}\label{eq:solFPangle}
   \mathcal{P}(\phi,t-t^{\prime}\vert\phi^{\prime})&=&\frac {1}{2\pi} \sum_{n=-\infty}^{\infty} e^{in(\phi-\phi^{\prime})}e^{- n^2 \overline{\Omega}(t-t^{\prime})},\\
   & = & \frac{1}{2\pi} \left(1+2\sum_{n=1}^{\infty} \cos[n(\phi-\phi^{\prime})] e^{-n^2 \overline{\Omega}(t-t^{\prime})} \right),\nonumber
\end{eqnarray}
where $\overline{\Omega}(t-t^{\prime})=\int_{0}^{t-t^{\prime}} ds\, \Omega(s)$ for an arbitrary autocorrelation function of the rotational noise $\xi_R(t)$. Eq.~(\ref{eq:solFPangle}) is the well-known Gaussian distribution \emph{wrapped} around the circle, and can be written in terms of the Jacobi theta function $\vartheta_{3}(z,q)=\sum_{n=-\infty}^{\infty}q^{n^{2}}e^{2inz}=1+2\sum_{n=1}^{\infty}q^{n^{2}}\cos2nz$ \cite{abramowitzBook}, i.e.,
\begin{equation}
    \mathcal{P}\left(\phi,t\vert\phi^{\prime}\right)= \frac{1}{2\pi}\vartheta_{3}\left(\frac{\phi-\phi^{\prime}}{2},e^{-\overline{\Omega}(t)}\right).
\end{equation}
By use of the Poisson summation formula \cite{guinand1941,applebaum2016}, $\mathcal{P}\left(\phi,t\vert\phi^{\prime}\right)$ can be rewritten as the equivalent Gaussian distribution wrapped around the circle
\begin{eqnarray}
\mathcal{P}\left(\phi,t\vert\phi^{\prime}\right)&=&{\sqrt{\frac{1}{4\pi \overline{\Omega}(t)}}\sum_{n=-\infty}^{\infty}\exp\left\{-\frac{[2\pi n-(\phi-\phi^{\prime})]^{2}}{4\overline{\Omega}(t)}\right\},}\nonumber\\
&=&\sqrt{\frac{1}{4\pi \overline{\Omega}(t)}}\exp\left[-\frac{(\phi-\phi^{\prime})^{2}}{4\overline{\Omega}(t)}\right]   \vartheta_{3}\left(\frac{\pi(\phi-\phi^{\prime})}{2 i\overline{\Omega}(t)},e^{-\frac{\pi^2}{\overline{\Omega}(t)}}\right),
\end{eqnarray}
from which the Gaussian distribution appears explicitly as a factor.

Notice that, in a short time interval $t$, $\mathcal{P}(\phi,t\vert\phi^{\prime})$ peaks sharply around $\phi^{\prime}$, meaning that the transition from the direction of motion $\phi^{\prime}$ to the new one $\phi$, occurs more frequently in the forward direction, i.e., around the direction of motion $\phi^{\prime}$. As the duration $t$ of the time interval of the transition becomes larger, the peak is smoothed out, thus converging to a uniform transition distribution in the asymptotic limit $t\rightarrow\infty$. The mean value of quantities of the form $f(\phi-\phi^{\prime})$, defined by 
\begin{equation}\label{eq:meantrans}
\langle f(\phi-\phi^{\prime})\rangle_{t}=\int_{0}^{2\pi}d\phi\int_{0}^{2\pi}d\phi^{\prime}f(\phi-\phi^{\prime})\mathcal{P}(\phi,t\vert\phi^{\prime})P(\phi^{\prime},0), 
\end{equation}
is of special interest. In particular, it can be shown from (\ref{eq:solFPangle}) and (\ref{eq:meantrans}) that
\begin{equation}
    \left\langle e^{in(\phi-\phi^{\prime})}\right\rangle_{t}=e^{-n^{2}\overline{\Omega}(t)}
\end{equation}
or equivalently
\numparts
\begin{eqnarray}
 \left\langle \cos n\left[\phi-\phi^{\prime}\right]\right\rangle_{t}&=e^{-n^{2}\overline{\Omega}(t)},\\
 \left\langle \sin n\left[\phi-\phi^{\prime}\right]\right\rangle_{t}&=0,
\end{eqnarray}
\endnumparts
which give the contributions to the moment expansion of $\mathcal{P}(\phi,t\vert\phi^{\prime})$ when this is written as
\begin{equation}
\mathcal{P}(\phi,t\vert\phi^{\prime})=\frac{1}{2\pi}\sum_{n=-\infty}^{\infty}e^{in(\phi-\phi^{\prime})}\left\langle e^{in(\phi-\phi^{\prime})}\right\rangle_{t}.
\end{equation}

The probability density $P(\phi,t)$, independent of the initial angle $\phi^{\prime}$, is obtained from $\mathcal{P}(\phi,t\vert\phi^{\prime})$ as
\begin{equation}
 P(\phi,t)=\int_{0}^{2\pi}d\phi^{\prime}\mathcal{P}(\phi,t\vert\phi^{\prime})P(\phi^{\prime},0)   
\end{equation}
where $P(\phi^{\prime},0)$ denotes the initial distribution of the particle direction of motion.

In the case of fractional Gaussian noise with autocorrelation given by Eq.~(\ref{fGnoise}), Eq.~(\ref{eq:intcorrfGnoise}) leads to
\begin{equation}\label{eq:intintcorrfGnoise}
\overline{\Omega}(t) = D_Ht^{2H},
\end{equation}
which corresponds to half the variance of the fluctuations of $\varphi$, see Eq.~(\ref{msdfGmotion}), thus yielding 
\begin{eqnarray}\label{eq:angdistr}
 P(\phi,t) & = & \frac {1}{2\pi} \vartheta_3 \left(\frac{\phi}{2},e^{-D_Ht^{2H}}\right),\nonumber\\
 & = & \frac{e^{ -\frac{\phi^2}{4D_H t^{2H}}}}{\sqrt{4\pi D_H t^{2H}}}\vartheta_3 \left( \frac{\pi \phi}{2iD_H t^{2H}}, e^{-\frac{\pi^2}{D_Ht^{2H}}} \right),
\end{eqnarray}
for the initial angular distribution $P(\phi^{\prime},0)=\delta(\phi^{\prime})$.
The angular probability density given by Eq.~(\ref{eq:angdistr}) retains a Gaussian-like shape at sufficiently short time scales $\bigl(t \ll D_H^{-\frac{1}{2H}}\bigr)$, then it spreads over the entire interval $0 \le \phi < 2\pi$ as $t$ increases, i.e., it is wrapped around the circle, and converges to the uniform distribution $P(\phi,t) = \frac{1}{2\pi}$ at $t \gg D_H^{-\frac{1}{2H}}$ for all $0 < H < 1$. However, depending on the specific value of $H$, different profiles of $P(\phi,t)$ are observed at a given time $t > 0$, for the same initial condition $P(\phi,0) = \delta(\phi)$. This is illustrated in Figs.~\ref{fig:pdfangle}(a),~\ref{fig:pdfangle}(b) and~\ref{fig:pdfangle}(c) where we plot the angular density $P(\phi,t)$ for different values of the Hurst exponent, $H = 0.1$ (antipersistent orientational dynamics), 0.5 (Brownian orientational dynamics), and 0.9 (persistent orientational dynamics), respectively, at different times $t = 0.01D_H^{-\frac{1}{2H}},0.1D_H^{-\frac{1}{2H}},D_H^{-\frac{1}{2H}},10D_H^{-\frac{1}{2H}},100D_H^{-\frac{1}{2H}}$.

For antipersistent rotational noise [$H < \frac{1}{2}$, see Fig.~\ref{fig:pdfangle}(a) for $H=0.1$], $P(\phi,t)$ broadens quickly over the full angular domain $(-\pi,\pi]$ during $0 < t < D_H^{-\frac{1}{2H}}$ but markedly preserves a peak around $\phi=0$ [see thick solid, dotted and dashed lines in Fig.~\ref{fig:pdfangle}(a) for $t = 0.01D_H^{-\frac{1}{2H}},0.1D_H^{-\frac{1}{2H}}$, and $D_H^{-\frac{1}{2H}}$ respectively]. This indicates that in this time regime, the particle direction of motion scatters more frequently in the forward direction even when any change of the orientation is probable (\emph{rectification of motion}), thereby causing a highly correlated motion. In contrast, $P(\phi,t)$ converges very slowly to the uniform angular density $(2\pi)^{-1}$ for time intervals $t > D_H^{-\frac{1}{2H}}$, retaining a smooth peak at $\phi=0$ [see Fig.~\ref{fig:pdfangle}(a) for $H = 0.1$ at $t = 10D_H^{-\frac{1}{2H}},100D_H^{-\frac{1}{2H}}$, dotted-dashed and thin solid lines, respectively], thus leading to a strong persistence of translational motion, as shown in Fig.~\ref{fig:trajectories}(c). The opposite trend is observed for persistent fractional noise ($H > \frac{1}{2}$). For instance, in Fig.~\ref{fig:pdfangle}(c) we show that, for $H = 0.9$, the initial delta peak $\delta(\phi)$ at $t = 0$ broadens rather slowly during $0 < t < D_H^{-\frac{1}{2H}}$. As a result, the particle  retains its direction of motion, whose persistence causes the translational looped trajectories shown in Fig. \ref{fig:trajectories}(g)-(h). On the other hand, a very fast convergence to the steady-state uniform value $ (2\pi)^{-1}$ occurs for time intervals $t > D_H^{-\frac{1}{2H}}$. For such a large value of $H$, the typical time scale needed to observe such a convergence is $t \sim 10 D_H^{-\frac{1}{2H}}$, see dotted-dashed line in Fig.~\ref{fig:pdfangle}(c). Only for the specific time interval $t = D_H^{-\frac{1}{2H}}$, the angular density profile is the same for all values of the Hurst exponent, and is given by $P\left(\phi,D_H^{-\frac{1}{2H}}\right) = \frac{1}{2\pi} \vartheta_3 \left( \frac{\phi}{2},e^{-1} \right)$, see dashed lines in Figs.~\ref{fig:pdfangle}(a)-(c). 

We point out that the convergence of $P(\phi,t)$ to a uniform angular distribution suggests that an active particle with fractional rotational Brownian motion must exhibit active, but normal, diffusion at sufficiently long time scales, for both persistent and antipersistent rotational noise, as explicitly shown in Section~\ref{Sect:AD}. 

\section{Active diffusion}\label{Sect:AD}

\subsection{Velocity autocorrelation function}

\begin{figure}[t]
\includegraphics[width=0.8\textwidth]{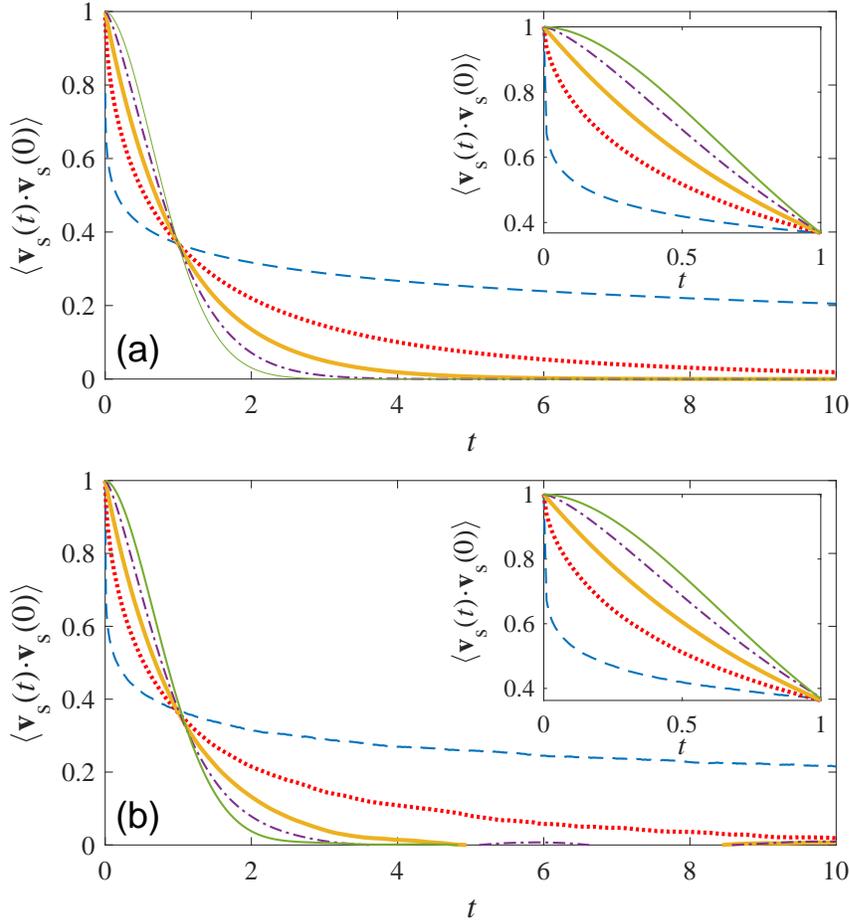}
 \caption{(a) Velocity autocorrelation function computed from Equation (\ref{eq:velcorr}), for different values of $H$: $0.1$ (dashed line), $0.3$ (dotted line), $0.5$ (thick solid line), $0.7$ (dotted-dashed line), and $0.9$ (thin solid line). Inset: expanded view for $0 \le t \le 1$. (b) Velocity autocorrelation function computed from simulated active trajectories with fractional rotational Brownian motion for the same values of $H$ as in (a), plotted with same line style.}\label{fig:velautocorr}
\end{figure}

We now compute the autocorrelation function of the swimming velocity, i.e., $\langle \boldsymbol{v}_{s}(s) \cdot \boldsymbol{v}_{s}(s') \rangle = v_0^2 \langle \hat{\boldsymbol{v}}(s) \cdot \hat{\boldsymbol{v}}(s')\rangle$, where the orientational correlation function can be expressed in terms of the angular coordinate $\phi$ as
\begin{equation}\label{eq:orientcorr}
\langle \hat{\boldsymbol{v}}(s) \cdot \hat{\boldsymbol{v}}(s')\rangle = \left\langle \cos [ \phi(s) - \phi(s')] \right\rangle,
\end{equation}
which is equivalent to $\langle \cos [\phi - \phi'] \rangle_{s-s^{\prime}}$ for $s\ge s^{\prime}$.
Therefore, Eq.~(\ref{eq:orientcorr}) can be explicitly computed by means of 
\begin{equation}\label{eq:angcorr}
\langle \hat{\boldsymbol{v}}(s) \cdot \hat{\boldsymbol{v}}(s')\rangle  = \int_0^{2\pi} \int_0^{2\pi} d\phi \, d\phi'\, \cos(\phi - \phi')\mathcal{P}(\phi,s-s'|\phi')P(\phi',s'),
\end{equation}
where $\mathcal{P}(\phi,s-s'|\phi')$ is the transition probability density from $\phi'$ at time $s'$ to $\phi$ at time $s$, as was introduced in Sect. \ref{subsect:angPDF}, whereas $P(\phi',s')$ is the angular probability density at time $s' \ge 0$ given by (\ref{eq:angdistr}). For $s \ge s' \gg D_H^{-\frac{1}{2H}}$, $\langle \hat{\boldsymbol{v}}(s) \cdot \hat{\boldsymbol{v}}(s')\rangle$ becomes stationary, where $P(\phi',s') \rightarrow (2\pi)^{-1}$, while
$\mathcal{P}(\phi,s-s'|\phi') = P(\phi-\phi',s-s')$. Using the expressions given in Eqs.~(\ref{eq:solFPangle}), (\ref{eq:intintcorrfGnoise}) and~(\ref{eq:angcorr}), we find that the velocity autocorrelation function is given explicitly by
\begin{equation}\label{eq:velcorr}
    \langle \boldsymbol{v}_{s}(s) \cdot \boldsymbol{v}_{s}(s') \rangle = v_0^2 \exp \left[-D_H (s-s')^{2H} \right].
\end{equation}
Eq.~(\ref{eq:velcorr}) corresponds to: a \emph{stretched exponential} when $0 < H < \frac{1}{2}$ describing a highly correlated motion; a \emph{pure exponential} if $H = \frac{1}{2}$ describing Brownian correlations of the direction of motion; and a \emph{compressed exponential} when $\frac{1}{2} < H < 1$ that describes short-ranged correlations of the direction of motion. In Fig.~\ref{fig:velautocorr}(a) we plot the autocorrelation function of the swimming velocity given by Eq.~(\ref{eq:velcorr}), $\langle \boldsymbol{v}_{s}(t) \cdot \boldsymbol{v}_{s}(0) \rangle = \langle \boldsymbol{v}_{s}(s'+t) \cdot \boldsymbol{v}_{s}(s') \rangle$, as a function of the time lag $t = s - s'$ for different values of $H$. We check that they perfectly agree with the numerical results shown in Fig.~\ref{fig:velautocorr}(b). In addition, we find that, regardless of $H$, the velocity autocorrelation attains the value $v_0^2e^{-1}$ at $t = D_H^{-\frac{1}{2H}}$. Nevertheless, for other values of $t$, different regimes are observed depending on $H$. For instance, for $0 < H < \frac{1}{2}$, $\langle \boldsymbol{v}_{s}(t) \cdot  \boldsymbol{v}_{s}(0) \rangle$ decays sharply from $v_0^2$ to $v_0^2 e^{-1}$ for $0 \le t < D_H^{-\frac{1}{2H}}$, as highlighted in the insets of Figs.~\ref{fig:velautocorr}(a) and \ref{fig:velautocorr}(b), followed by a very slow decrease for $t \ge D_H^{-\frac{1}{2H}}$.
On the other hand, for $H=\frac{1}{2}$ we find that a purely exponential decay is recovered, i.e., $\langle \boldsymbol{v}_{s}(t) \cdot \boldsymbol{v}_{s}(0) \rangle = v_0^2 \exp(-D_{1/2} t)$, for which the particle orientation is driven by Gaussian white noise with rotational diffusion coefficient $D_{1/2}$ and decorrelation time set by $\tau_{1/2} = D_{1/2}^{-1}$. 
Finally, for $\frac{1}{2} < H < 1$ (persistent rotational noise) the velocity autocorrelation function decreases more slowly in time for $0 \le t < D_H^{-\frac{1}{2H}}$, whereas it quickly goes to 0 for $t \ge D_H^{-\frac{1}{2H}}$. In particular, as $H \rightarrow 1$, the velocity autocorrelation approaches a Gaussian decay, i.e., $v_0^2\exp(-D_1 t^2)$. Consequently, for $\frac{1}{2} \le  H < 1$, the typical decorrelation time of the swimming velocity is $\lesssim D_H^{-\frac{1}{2H}}$, whereas for $0 < H < \frac{1}{2}$, long-range temporal correlations of the particle orientation lead to a rather high persistence of the swimming velocity.

\begin{figure*}
\includegraphics[width=0.9\textwidth]{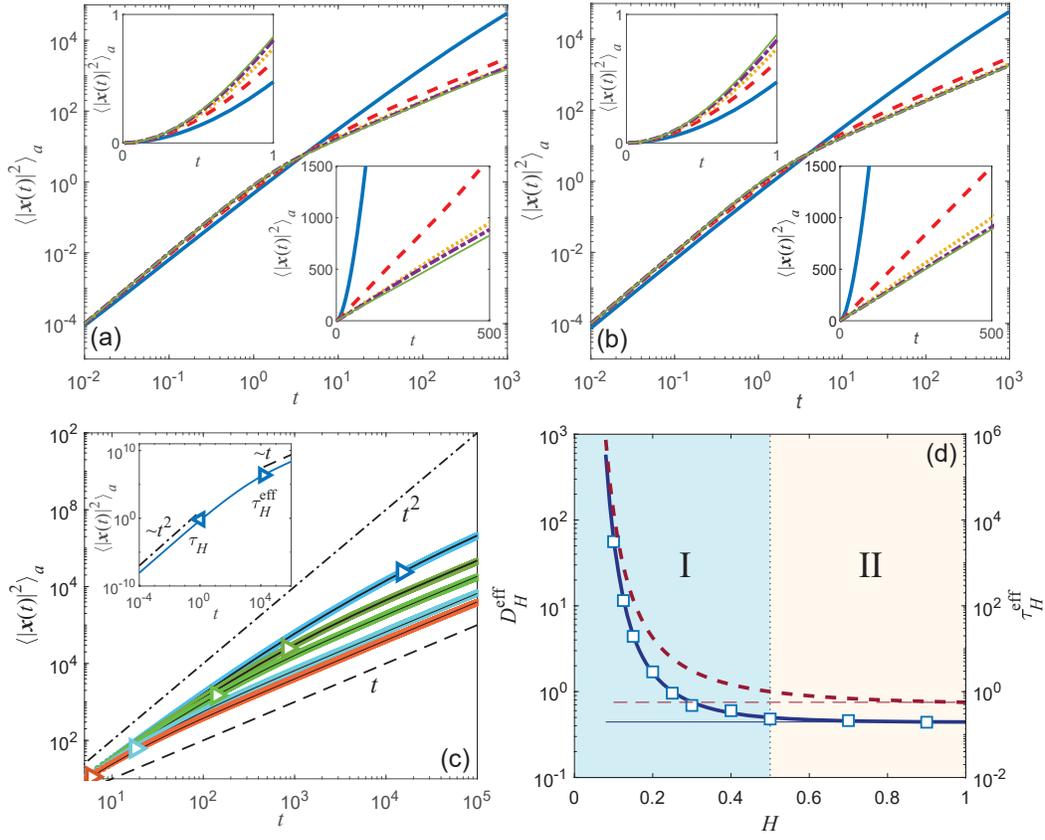}
\caption{(a) Translational mean-squared displacement given by Eq. (\ref{eq:msdtrans}), for different values of the Hurst exponent: $H
= 0.1$ (thick solid line),
$H = 0.3$ (dashed line),
$H = 0.5$ (dotted line),
$H = 0.7$ (dashed-dotted line), and $H = 0.9$ (thin solid
line).  The top-left and bottom-right insets show a linear-linear representation of the main plot at short and long time scales, respectively. (b) Translational mean-squared displacements obtained from simulated trajectories. Same color code and same
line style as in Fig. \ref{fig:MSD2dtrans}(a). (c) Long-time behavior of the translational mean square displacements for an active particle driven by antipersistent rotational fractional noise. From top to botton; $H = 0.1, 0.125, 0.15, 0.2, 0.25$. The colored symbols correspond to the curves obtained from numerical simulations, whereas the black solid lines represent the analytical expression given by Eq.~(\ref{eq:msdtrans}). The triangles ($\triangleright$) depict the corresponding location of the effective rotational time,
$\tau_H^{\mathrm{eff}}$, above which active diffusion emerges. Inset: translational mean-squared displacement for $H=0.1$. The triangles ($\triangleleft$) and ($\triangleright$) indicate the location of the persistence time $\tau_H$ and the effective rotational time $\tau_H^{\mathrm{eff}}$, respectively, which define the time interval $[\tau_H,\tau_H^{\mathrm{eff}}]$ in which anomalous diffusion is observed. (d) Active diffusion coefficient (solid line, left axis) and effective rotational diffusion time (dashed line, right axis), as a function of $H$. The horizontal solid and dashed lines represent the limit values as $H \rightarrow 1$, $D^{\mathrm{eff}}_H = \frac{\sqrt{\pi}}{4}$ and $\tau_H^{\mathrm{eff}} = \frac{1}{\sqrt{\pi}}$, respectively, whereas the squares are numerical values of $D_H^{\mathrm{eff}}$ computed from simulations. The vertical dotted line separates the two distinct regimes of active motion: I) long-range correlations of the swimming velocity, and II) fast decay of the swimming-velocity autocorrelation function.}
\label{fig:MSD2dtrans}
\end{figure*}

\subsection{Mean-squared displacement}

The translational mean-squared displacement of the particle position, $\boldsymbol{x} = (x,y)$, can determined by employing the relation 
\begin{eqnarray}
\langle |{\boldsymbol{x}}(t)|^2 \rangle & = & \int_0^t \int_0^t ds \, ds' \left\langle \frac{d}{ds}\boldsymbol{x}(s) \cdot \frac{d}{ds'}\boldsymbol{x}(s') \right\rangle,  \label{eq:msd2Dtrans} \label{eq:transmsd2D} \nonumber\\
& = & \int_0^t \int_0^t ds \, ds' \langle \boldsymbol{\xi}_{T}(s) \cdot \boldsymbol{\xi}_{T}(s') \rangle
  + \int_0^t \int_0^t ds \, ds' \langle \boldsymbol{v}_{s}(s) \cdot \boldsymbol{v}_{s}(s') \rangle. \label{eq:transmsd2Dcomponents}
\end{eqnarray} 
The first term on the right hand-side of Eq.~(\ref{eq:transmsd2Dcomponents}), which will be denoted by $\langle |{\boldsymbol{x}}(t)|^2 \rangle_p$, represents the \emph{passive} component of the mean-squared displacement due to translational velocity fluctuations, $\boldsymbol{\xi}_T(t)$. Since we assume that $\boldsymbol{\xi}_T(t)$ is delta-correlated in the model (\ref{modelo}), this yields trivially the diffusive contribution 
\begin{equation}\label{eq:passmsdtrans}
   \langle |{\boldsymbol{x}}(t)|^2 \rangle_p = 4D_T t.
\end{equation}
On the other hand, the second term on the right hand-side of Eq.~(\ref{eq:transmsd2Dcomponents}), which will be denoted by $\langle |{\boldsymbol{x}}(t)|^2 \rangle_a$, originates from the orientational changes in the swimming velocity driven by fractional rotational Brownian noise and can be rewritten as
\begin{equation}\label{eq:transmsd2Dact}
\langle |{\boldsymbol{x}}(t)|^2 \rangle_a = 2v_{0}^{2}\int_0^t ds\int_0^s  ds' \langle\cos(\phi-\phi^{\prime})\rangle_{s-s^{\prime}}.
\end{equation}
Henceforth, we focus on the nontrivial \emph{active} contribution to the translational mean-squared displacement, i.e., $\langle |{\boldsymbol{x}}(t)|^2 \rangle_a = \langle |{\boldsymbol{x}}(t)|^2 \rangle - \langle |{\boldsymbol{x}}(t)|^2 \rangle_p$. Thus, from Eqs.~(\ref{eq:velcorr}) and~(\ref{eq:transmsd2Dact}), we can derive in a straightforward manner the general expression for this active component for all $0 < H < 1$, namely
\begin{eqnarray}
  \langle |{\boldsymbol{x}}(t)|^2 \rangle_a &= &2v_{0}^{2}\int_0^t ds\int_0^s  ds'e^{-D_{H}{s^{\prime}}^{2H}}\nonumber\\
  &= &v_0^2 t^2 \sum_{k = 0}^{\infty} \frac{\bigl(-D_H t^{2H}\bigr)^k}{k!(1+k H)(1+2k H)}\label{eq:msdtrans1}.
\end{eqnarray}
Eq.~(\ref{eq:msdtrans1}) can be rewritten as
\begin{equation}\label{eq:msdtrans}
   \langle |{\boldsymbol{x}}(t)|^2 \rangle_a  =  \frac{v_0^2}{H D_H^{\frac{1}{H}}}\left[ \gamma\left(\frac{1}{2H},D_H t^{2H}\right)D_H^{\frac{1}{2H}}t 
    - \gamma\left(\frac{1}{H},D_Ht^{2H}\right) \right],
\end{equation}
where $\gamma(\nu,z) = \int_0^z t^{\nu - 1} e^{-t} dt$ is the lower incomplete gamma function~\cite{abramowitzBook}. 
In Figs.~\ref{fig:MSD2dtrans}(a) and~\ref{fig:MSD2dtrans}(b), we plot some exemplary mean-squared displacements for different $H$, computed by means of Eq.~(\ref{eq:msdtrans}) and from the simulated trajectories, respectively. We verify that the analytic expression (\ref{eq:msdtrans}) for arbitrary $H$ agrees very well with the numerical results.

An apparent expression can be readily derived from Eq.~(\ref{eq:msdtrans1}) for the particular case $H = \frac{1}{2n}$, where $n = 1,2,\ldots$. In such a case, the mean-squared displacement can be expressed as the following finite sum
\begin{eqnarray}\label{eq:msdtrans2n}
    \langle |{\boldsymbol{x}}(t)|^2 \rangle_a & = & \frac{2n v_0^2}{D_{{1}/{2n}}^{2n}}\left\{ (n-1)!\left[ 1 - e^{-D_{{1}/{2n}}t^{\frac{1}{n}}} \sum_{k=0}^{n-1}\frac{D^k_{{1}/{2n}}t^{\frac{k}{n}}}{k!} \right] D_{{1}/{2n}}^n t \right. \nonumber\\
&&\left.+ (2n-1)!\left[e^{-D_{{1}/{2n}}t^{\frac{1}{n}}}\sum_{k=0}^{2n-1}\frac{D^k_{{1}/{2n}}t^{\frac{k}{n}}}{k!} -1 \right] \right\}.
\end{eqnarray}
For $n = 1$, i.e. $H = \frac{1}{2}$, we recover the well-known expression of persistent Brownian motion with rotational diffusion coefficient $D_{1/2}$
\begin{equation}\label{eq:msdABM}
 \langle |{\boldsymbol{x}}(t)|^2 \rangle_a =\frac{2v_{0}^{2}}{D_{1/2}^{2}}\left(D_{1/2}t+e^{-D_{1/2}t}-1\right),
\end{equation}
whereas for $n = 2$, i.e., $H = \frac{1}{4}$ (antipersistent rotational noise), we get
\begin{equation}\label{eq:msdtransH0_25}
   \langle |{\boldsymbol{x}}(t)|^2 \rangle_a =   \frac{4v_0^2}{D_{1/4}^{4}}\left[2e^{-D_{1/4}\sqrt{t}}\left( 3 + 3D_{1/4} \sqrt{t} + D_{1/4}^2 t \right)+ D_{1/4}^2 t - 6 \right].
\end{equation}
From Eq.~(\ref{eq:msdtrans1}), it can be easily seen that in the limit of fully antipersistent rotational noise, $H \rightarrow 0$, the mean-squared displacement is ballistic at all times, i.e., 
\begin{equation}\label{eq:fullpersist}
\langle |{\boldsymbol{x}}(t)|^2 \rangle_a \rightarrow \frac{1}{e} v_0^2 t^2. 
\end{equation}
Eq.~(\ref{eq:fullpersist}) represents the limit of infinite persistence of self-propelled motion, characterized by a constant value of the velocity autocorrelation function~$ \langle \boldsymbol{v}_{s}(s) \cdot \boldsymbol{v}_{s}(s') \rangle \rightarrow v_0^2 / e$ for all $s > s'$, and an effective swimming speed $v_0/\sqrt{e}$. 
The other extreme limit corresponds to fully persistent rotational fractional Brownian noise, $H \rightarrow 1$, for which we find
\begin{equation}\label{eq:msdtransH1}
    \langle |{\boldsymbol{x}}(t)|^2 \rangle_a  \rightarrow  \frac{v_0^2}{D_1}\left[ \sqrt{\pi}\, {\mathrm{erf}}\left({D_1}^{\frac{1}{2}} t\right) D_1^{\frac{1}{2}}t  
     + e^{-D_1 t^2} -1 \right],
\end{equation}
where $\mathrm{erf}(z) = \frac{2}{\sqrt{\pi}}\int_{0}^{z} dt e^{-t^2}$ is the error function. Note that, in this case, the velocity autocorrelation approaches the Gaussian decay $\langle \boldsymbol{v}_{s}(s) \cdot \boldsymbol{v}_{s}(s') \rangle \rightarrow v_0^2\exp \left[-D_1 (s-s')^2 \right]$. In the long-time regime $D_{1}^{1/2}t\gg1$, ${\mathrm{erf}}({D_1}^{\frac{1}{2}} t)\approx1$ and hence the linear dependence $\sqrt{\pi/D_{1}}v_{0}^{2}\,t$ is obtained.

For all values of $0 < H < 1$, two important limiting cases are observed. First, since $\gamma(\nu,z) \rightarrow \nu^{-1}z^{\nu}$ as $z \rightarrow 0$, then for $t \ll D_H^{-\frac{1}{2H}}$
\begin{equation}\label{eq;shortmsd}
 \langle |{\boldsymbol{x}}(t)|^2 \rangle_a \approx v_0^2 t^2. 
\end{equation}
This limit corresponds to the characteristic ballistic behavior, which is expected to happen due to the persistence of the swimming velocity ${\boldsymbol{v}}_s(t)$ at sufficiently short time scales. However, at intermediate time scales two qualitatively distinct regimes can be distinguished depending on the behavior of the mean-squared displacement with respect to the value of $H$:
\begin{itemize}
\item[I)] First, for $H < 0 < \frac{1}{2}$ the short-time ballistic regime is rapidly hindered at $t \lesssim D_H^{-\frac{1}{2H}}$ by the antipersistence of the rotational noise. This results in an intermediate \emph{anomalous} regime where $\langle |{\boldsymbol{x}}(t)|^2 \rangle_a$ grows with time $t$ slower than $\sim t^2$ but faster than $\sim t$ over a broad temporal interval up to several times $D_H^{-\frac{1}{2H}}$, as shown in Figs.~\ref{fig:MSD2dtrans}(a) and~\ref{fig:MSD2dtrans}(b) for $H = 0.1,0.3$, and in Fig.~\ref{fig:MSD2dtrans}(c) for $H = 0.1,0.125,0.15,0.2$. This is also consistent with the long-range temporal correlations of the swimming velocity, which persist even for time scales comparatively larger than $D_H^{-\frac{1}{2H}}$, as shown in Fig.~\ref{fig:velautocorr}. 
\item[II)] On the other hand, for $\frac{1}{2}\le H <1$, the persistence of the rotational noise allows to fully preserve the ballistic behavior up to time scales $t \approx D_H^{-\frac{1}{2H}}$, see Figs.~\ref{fig:MSD2dtrans}(a) and~\ref{fig:MSD2dtrans}(b) for $H =0.5,0.7,0.9$. For $t \gtrsim D_H^{-\frac{1}{2H}}$, $\langle |{\boldsymbol{x}}(t)|^2 \rangle_a$ reaches quickly a diffusive behavior, caused by the complete decorrelation of the particle orientation, as verified in Fig.~\ref{fig:velautocorr}. Note that in this case, the resulting slope of linear behavior of the mean-squared displacement varies very weakly with $H$, as observed in the bottom-right insets of Figs.~\ref{fig:MSD2dtrans}(a) and~\ref{fig:MSD2dtrans}(b).
\end{itemize}
Furthermore, the second important limit is obtained at sufficiently long-time scales ($t \gg D_H^{-\frac{1}{2H}}$), for which we find 
\begin{equation}\label{eq:actdiff}
    \langle |{\boldsymbol{x}}(t)|^2 \rangle_a \approx \frac{v_0^2}{H D_H^{\frac{1}{2H}}}\Gamma\left( \frac{1}{2H} \right) t,
\end{equation}
where $\Gamma(\nu) = \int_0^{\infty} t^{\nu - 1} e^{-t}dt$ is the complete gamma function. Remarkably, Eq.~(\ref{eq:actdiff}) reveals that active diffusion emerges in the long-time limit for all values of the Hurst exponent, $0 < H < 1$: $\langle |{\boldsymbol{x}}(t)|^2 \rangle_a \approx 4D_H^{\mathrm{eff}} t$, where the resulting  active diffusion coefficient is
\begin{eqnarray}\label{eq:effdiffcoeff}
    D^{\mathrm{eff}}_H & = & \frac{v_0^2}{4 H D_H^{\frac{1}{2H}}}\Gamma\left( \frac{1}{2H} \right),\nonumber\\
    & = & \frac{1}{4H}  \Gamma\left( \frac{1}{2H} \right) \mathcal{D}_H.
\end{eqnarray}
Indeed, in Fig.~\ref{fig:MSD2dtrans}(c), we show that, even for regime I ($0<H<\frac{1}{2}$), for which an anomalous growth of the mean-squared displacement occurs at intermediate time scales, a diffusive behavior is reached at sufficiently long time scales. In such a case, the slope of the long-time linear behavior of $\langle |{\boldsymbol{x}}(t)|^2 \rangle_a$ becomes very sensitive to small variations of the Hurst exponent: the smaller the value of $H$, the larger the resulting active diffusion coefficient, as illustrated in Fig.~\ref{fig:MSD2dtrans}(c). By performing a linear fit of the long-time behavior of mean-squared displacements obtained from the simulated trajectories, we compute the numerical values of $D_H^{\mathrm{eff}}$, which are plotted as squares in Fig. ~\ref{fig:MSD2dtrans}(d). For comparison, we also plot as a solid line the dependence of $D_H^{\mathrm{eff}}$ on $H$ given by Eq. (\ref{eq:effdiffcoeff}), thereby showing a very good agreement with the numerical results. Once again, two distinct behaviors of $D^{\mathrm{eff}}_H$ are observed as a function of $H$, which coincide with the existence of the two different regimes (I and II) previously identified. For regime I, the active diffusion coefficient exhibits a sharp monotonic increase with decreasing $H$, and
diverges as $H \rightarrow 0$. In addition, it approaches the value $D_{1/2}^{\mathrm{eff}} = \frac{1}{2} \mathcal{D}_{1/2} = \frac{1}{2}v_0^2 D_{1/2}^{-1}$ as $H \rightarrow \frac{1}{2}$. On the other hand, for regime II, $D^{\mathrm{eff}}_H$ varies very weakly with $H$: starting from $D_{1/2}^{\mathrm{eff}}$ it decreases monotonically as $H$ increases and converges to the value $D^{\mathrm{eff}}_1 = \frac{\sqrt{\pi}}{4} \mathcal{D}_{1} =  \frac{v_0^2}{4}\sqrt{\frac{\pi}{D_1}}$ as $H \rightarrow 1$.

The previous results suggest that two relevant time scales are necessary to describe active motion driven by fractional rotational Brownian noise. The first is the natural time scale
\begin{equation}\label{eq:persisttime}
\tau_H \equiv D_H^{-\frac{1}{2H}},
\end{equation}
which represents a \emph{persistence time} over which the active particle is able to keep on average a constant swimming velocity despite the angular fluctuations.  On the contrary, a second time scale, which will be denoted by $\tau_H^{\mathrm{eff}}$, represents the time needed for the particle orientation to become completely decorrelated and uniformly distributed over $[0,2\pi)$. Therefore, $\tau_H^{\mathrm{eff}}$ can be interpreted as an \emph{effective rotational time}, similar to $\tau_{1/2} = D_{1/2}^{-1}$ defined for active Brownian motion ($H=0.5$) as the time scale at which the autocorrelation function of the particle orientation decays to $1/e$. In fact, for this particular value of the Hurst exponent, both time scales coincide: $\tau_{1/2} = \tau_{1/2}^{\mathrm{eff}}$. 
However, for $H \neq 1/2$, it is expected that $\tau_{H}^{\mathrm{eff}}$ could be different from $\tau_H$ due to the  non-exponential decay of the velocity autocorrelation function. In order to determine $\tau_H^{\mathrm{eff}}$, we realize that a diffusive behavior of $\langle |{\boldsymbol{x}}(t)|^2 \rangle_a$ must be observed for $t \gtrsim \tau_H^{\mathrm{eff}}$. Thus, taking into account that $\gamma(\nu,z) \rightarrow \Gamma(\nu)$ as $z \rightarrow  \infty$, by applying the condition $\Gamma \left(\frac{1}{2H} \right) D_H^{\frac{1}{2H}}t  \gg \Gamma\left( \frac{1}{H} \right)$ to Eq.~(\ref{eq:msdtrans}) we find
\begin{eqnarray}\label{eq:rottime}
\tau_H^{\mathrm{eff}} & = &       
    \frac{\Gamma\left(\frac{1}{H}\right)}{\Gamma     \left( \frac{1}{2H} \right)} D_H^{-\frac{1}{2H}},\nonumber\\
    & = &  \frac{\Gamma\left(\frac{1}{H}\right)}{\Gamma     \left( \frac{1}{2H} \right)} \tau_H.
\end{eqnarray}
Indeed, for $H = \frac{1}{2}$, Eq.~(\ref{eq:rottime}) reduces to the well known expression $\tau_{1/2}^{\mathrm{eff}} = D_{1/2}^{-1} = \tau_{1/2}$ for pure rotational Brownian noise in two dimensions. In Fig.~\ref{fig:MSD2dtrans}(d) we show as a dashed line the dependence of $\tau_H^{\mathrm{eff}}$ on $H$ given by Eq.~(\ref{eq:rottime}). For regime I,  $\tau_H^{\mathrm{eff}}$ exhibits a very pronounced increase as $H$ decreases, and diverges as $H \rightarrow 0$. It should be noted that, in the case of antipersistent rotational noise, the increase of $\tau_H^{\mathrm{eff}}$ on $H$ is much more pronounced than that of $D_H^{\mathrm{eff}}$, as shown in Fig.~\ref{fig:MSD2dtrans}(d). For instance, for $H = 0.1$, $\tau_{0.1}^{\mathrm{eff}} = 15120 \,\tau_{0.1}$, whereas $D_{0.1}^{\mathrm{eff}} = 60 \,\mathcal{D}_{0.1}$. In Fig.~\ref{fig:MSD2dtrans}(c) we represent as triangles the location of $\tau_H^{\mathrm{eff}}$ on the mean-squared displacement curves , $\langle |{\boldsymbol{x}}(t)|^2 \rangle_a$ vs. $t$, for different $0 < H < \frac{1}{2}$. We verify that active diffusion emerges if the elapsed time $t$ is only slightly larger than the values of $\tau_H^{\mathrm{eff}}$ determined by means of Eq.~(\ref{eq:rottime}). Note that the separation between the persistence time and the effective rotational time opens a time interval $[\tau_H,\tau_H^{\mathrm{eff}}]$ over which the active motion is neither ballistic nor diffusive, as illustrated in the inset of Fig.~\ref{fig:MSD2dtrans}(c). The length of this interval where anomalous active motion occurs broadens as $H$ decreases, as illustrated in Fig.~\ref{fig:MSD2dtrans}(c) for different values of $0 < H < \frac{1}{2}$.

The opposite behavior is observed for regime II: with increasing $H$, $\tau_H^{\mathrm{eff}}$ decreases monotonically from the value  $\tau_{1/2}^{\mathrm{eff}} = D_{1/2}^{-1}$ at $H = 1/2$, thus becoming smaller than $\tau_H$. In this case, the dependence of  $\tau_H^{\mathrm{eff}}$ on $H$ is much less pronounced than in I, where the limiting value as $H \rightarrow 1$ is $\tau_1^{\mathrm{eff}} = \tau_{1}/\sqrt{\pi} = {1}/{\sqrt{\pi D_1}}$. Note that in this regime the time interval for the possible appearance of anomalous active motion, $[\tau_H^{\mathrm{eff}},\tau_H]$, is quite narrow.
In fact, the maximum relative difference between $\tau_H$ and $\tau_H^{\mathrm{eff}}$ is $(\tau_H - \tau_H^{\mathrm{eff}}) / \tau_H \approx 0.44$ as $H \rightarrow 1$. This implies that a rather abrupt transition from ballistic to active diffusion must occur in this regime at $t \approx D_H^{-\frac{1}{2H}}$, as verified in Figs.~\ref{fig:MSD2dtrans}(a) and~\ref{fig:MSD2dtrans}(b).

\section{Summary and final remarks}\label{Sect:Conc}

In this paper, we have investigated a two-dimensional model for a overdamped self-propelled particle moving at constant swimming speed, whose orientation is driven by fractional Brownian noise. The resulting dynamics of the swimming direction of the particle has deep consequences on its translational pattern of motion. Remarkably,  for positively correlated rotational noise, circular-like motion can be observed even in the absence of external elements that break the rotational symmetry, as found for active colloids swimming in viscoelastic media~\cite{narinder2018} or at large P\'eclet number~\cite{hu2019}. We have derived the corresponding Fokker-Planck equations, as well as the solution for the probability density function of the particle orientation for arbitrary values of the Hurst exponent $H$ of the fractional rotational noise. This in turn has allowed us to find analytical expressions for the swimming-velocity autocorrelation function and the translational mean-squared displacement, which reduce for $H = 0.5$ to the widely-known expressions of the conventional ABP model.

By analyzing the behavior of the derived quantities for different values of the Hurst exponent, we have identified two distinct regimes of active motion, marked by the influence of either the antipersistence or the persistence of the rotational noise. We have demonstrated that active diffusion effectively emerges in the asymptotic long-time limit regardless of the nature of the rotational noise. Moreover, we have provided an analytical expression for the active diffusion coefficient as a function of $H$, and checked that our results are in excellent agreement with numerical simulations of active trajectories evolving according to the proposed model. One remarkable finding of our work is the emergence of an $H$-dependent time scale which plays the role of an effective rotational-diffusive time, even though the orientational dynamics of the particle is not exponentially correlated if $H \neq 0.5$. The existence of such a time scale, in addition to the well-known persistence time, sets an interval over which the active motion exhibits anomalous diffusion. This is markedly apparent for antipersistent rotational noise with small Hurst exponent. In such a case, there exists a broad time interval characterized by long-range temporal correlations of the swimming velocity and an anomalous grow of the mean-squared displacement.

To our knowledge, our work is the first investigation of the effects of non-exponential orientational correlations in the motion of self-propelled particles. Thus, we expect that the results presented here will contribute to a better understanding of active motion in complex media with anomalous rotational diffusion, such as those found in many biological systems.  Further steps of our work could also address the effect of retarded memory effects in the rotational friction~\cite{sandev2014,rodriguez2015}, which could also modify the active diffusive behavior that emerges in the asymptotic limit. One more possible aspect to investigate is the influence of geometrical confinements, as it is known that rotational memory can significantly modify, e.g., the rectification of active particles in asymmetric periodic channels \cite{hu2017}.

\section*{Acknowledgements}
J.R.G.-S. acknowledges support from DGAPA-UNAM PAPIIT Grant No. IA103320. F.J.S. acknowledges support from DGAPA-UNAM PAPIIT-114717 and PAPIIT-IN110120.


\section*{References}

\begin{thebibliography}{99}
\bibitem{bechinger2016}  Bechinger C,  Di Leonardo R, L\"owen H, Reichhardt C, Volpe G and Volpe G 2016 {\it Rev. Mod. Phys.} \textbf{88}, 045006 

\bibitem{ramaswamy2010} Ramaswamy S 2010 {\it Annu. Rev. Condens. Matter Phys.} \textbf{1}, 323 

\bibitem{elgeti2015} Elgeti J, Winkler R G and Gompper G 2015 {\it Rep. Prog. Phys.} \textbf{78}, 056601 

\bibitem{taktikos2013} Taktikos J, Stark H and Zaburdaev V, 2013 {\it PLoS ONE} \textbf{8}, e81936 

\bibitem{darnton2007} Darnton N C and Berg H C, 2007 {\it Biophys. J.} \textbf{92}, 2230 

\bibitem{howse2007} Howse J R, Jones  R A L, Ryan A J, Gough T, Vafabakhsh R and Golestanian R 2007 {\it Phys. Rev. Lett.} \textbf{99}, 048102 

\bibitem{saragosti2012} Saragosti J, Silberzan P and Buguin A 2012
{\it PLoS One} \textbf{7}, e35412 

\bibitem{cates2013} Cates M E and Tailleur J 2013 {\it EPL} \textbf{101}, 20010 

\bibitem{tenhagen2011} Ten Hagen B,  van Teeffelen S and L\"owen H 2011 {\it J. Phys.: Condens. Matter} \textbf{23} 194119 

\bibitem{pototsky2012} Pototsky A and Stark H 2012 {\it EPL} \textbf{98}, 50004 

\bibitem{redner2013} Redner G S, Hagan M F and Baskaran A 2013 {\it Phys. Rev. Lett.} \textbf{110}, 055701 

\bibitem{bialke2013} Bialk\'e J,  L\"owen H, and Speck T 2013 {\it EPL} \textbf{103}, 30008 

\bibitem{sevilla2014} Sevilla F J and Gomez Nava L A 2014 {\it Physical Review E} \textbf{90}, 022130 

\bibitem{sevilla2015} Sevilla F J and Sandoval M 2015 {\it Phys. Rev. E} \textbf{91}, 052150 

\bibitem{basu2018} Basu U, Majumdar S N, Rosso A and Schehr G 2018 {\it Phys. Rev. E} \textbf{98}, 062121  

\bibitem{kurzthaler2018}Kurzthaler C, Devailly C, Arlt J, Franosch T, Poon  W C K, Martinez V A and Brown A T 2018 {\it Phys. Rev. Lett. \textbf{212}, 078001}

\bibitem{bregulla2015} Bregulla A P and Cichos F 2015 {\it Faraday Discuss.} \textbf{184}, 381 

\bibitem{gomezsolano2017} Gomez-Solano J R, Samin S, Lozano C, Ruedas-Batuecas P, van Roij R and Bechinger C 2017 {\it Sci. Rep.} \textbf{7}, 14891 

\bibitem{solon2015} Solon A P, Cates M E and Tailleur J 2015 {\it Eur. Phys. J. Spec. Top.} \textbf{224}, 1231 

\bibitem{vachier2019} Vachier J and Mazza M G 2019 {\it Eur. Phys. J. E} \textbf{42}, 11 

\bibitem{woillez2019} Woillez E, Zhao Y, Kafri Y, Lecomte V and Tailleur J, 2019 {\it Phys. Rev. Lett.} \textbf{122}, 258001 

\bibitem{tenhagenpre2011} Ten Hagen B, Wittkowski R and L\"owen  H 2011 {\it Phys. Rev. E }\textbf{84}, 031105 

\bibitem{zoettl2012} Z\"oettl A and Stark H 2012 {\it Phys. Rev. Lett.} \textbf{108}, 218104 

\bibitem{li2017} Li Y, Marchesoni F, Debnath T and Ghosh P K 2017 {\it Phys. Rev. E} \textbf{96}, 062138 

\bibitem{hu2017} Hu C-T, Wu J-C and Ai B-Q 2017 {\it J. Stat. Mech.} 053206.

\bibitem{duzgun2018} Duzgun A and Selinger J V 2018 {\it Phys. Rev. E} \textbf{97}, 032606 

\bibitem{wagner2019} Wagner C G, Hagan M H and Baskaran A 2019 {\it Phys. Rev. E} \textbf{100}, 042610 

\bibitem{caprini2019} Caprini L, Cecconi F and Marconi U M B 2019 {\it J. Chem. Phys.} \textbf{150}, 144903

\bibitem{wysocki2014} Wysocki A, Winkler R G and Gompper G 2014 {\it EPL} \textbf{105}, 48004


\bibitem{stenhammar2014} Stenhammar J, Marenduzzo D, Allen R J and Cates M E 2014 {\it Soft Matter} \textbf{10}, 1489 

\bibitem{richard2016} Richard D, L\"owen H and Speck T 2016 {\it Soft Matter} \textbf{12}, 5257 

\bibitem{speck2016} Speck T 2016 {\it EPL} \textbf{114} 30006 

\bibitem{pietzonka2016} Pietzonka P, Kleinbeck K and Seifert U 2016 {\it New J. Phys.} \textbf{18} 052001 

\bibitem{falasco2016} Falasco G, Pfaller R, Bregulla A P, Cichos F and Kroy K 2016 {\it Phys. Rev. E} \textbf{94}, 030602(R) 

\bibitem{gaspard2017} Gaspard P and Kapral R 2017 {\it J. Chem. Phys} \textbf{147} 211101 

\bibitem{shankar2018} Shankar S and Marchetti M C 2018 {\it Phys. Rev. E} 98, 020604(R) 

\bibitem{peruani2007} Peruani F and Morelli L G 2007 {\it Phys. Rev. Lett.} 99, 010602 

\bibitem{gosh2015} Ghosh P K, Li Y, Marchegiani G and Marchesoni F 2015 {\it J. Chem. Phys.} \textbf{143}, 211101 

\bibitem{debnath2016} Debnath D, Ghosh P K, Li Y, Marchesoni F and Li B 2016 {\it Soft Matter} \textbf{12}

\bibitem{narinder2018} Narinder N, Bechinger C and Gomez-Solano J R 2018 {\it Phys. Rev. Lett.} \textbf{121}, 078003 

\bibitem{sevilla2019} Sevilla F J, Rodríguez R F and Gomez-Solano J R 2019 {\it Phys. Rev. E} \textbf{100}, 032123 

\bibitem{gomezsolano2016} Gomez-Solano J R, Blokhuis A and Bechinger C 2016 {\it Phys. Rev. Lett.} \textbf{116}, 138301 

\bibitem{lozano2018} Lozano C, Gomez-Solano J R and Bechinger C 2018 {\it New J. Phys.} \textbf{20}, 015008 

\bibitem{lozano2019} Lozano C, Gomez-Solano J R and Bechinger C 2019 {\it Nat. Mater.} \textbf{18}, 1118 

\bibitem{saad2019} Saad S and Natale G 2019 {\it Soft Matter} \textbf{15}, 9909 

\bibitem{narinder2019} Narinder N, Gomez-Solano J R and Bechinger C 2019 {\it New J. Phys.} \textbf{21}, 093058 

\bibitem{chepizhko2013} Chepizhko O and Peruani F 2013 {\it Phys. Rev. Lett.} \textbf{111}, 160604 

\bibitem{tolic2004} Toli\'c-N{\o}rrelykke I M, Munteanu  E-L, Thon G, Oddershede L and Berg-S{\o}rensen K 2004 {\it Phys. Rev. Lett.} \textbf{93}, 078102 

\bibitem{wong2004} Wong I Y, Gardel M L, Reichman D R, Weeks E R, Valentine M T, Bausch A R and Weitz D A 2004 {\it Phys. Rev. Lett.} \textbf{92}, 178101 

\bibitem{jeon2013} Jeon  J-H, Leijnse N, Oddershede L B and Metzler R 2013 {\it New J. Phys.} \textbf{15} 045011 

\bibitem{thapa2019} Thapa S, Lukat N, Selhuber-Unkel C, Cherstvy A G and Metzler R 2019 {\it J. Chem. Phys.} \textbf{150}, 144901 

\bibitem{deschenes2001} Deschenes L A and Vanden Bout D A2001 {\it Science} \textbf{292}, 255 

\bibitem{cote2010} Cote Y, Senet P, Delarue P, Maisuradze G G and Scheraga H A 2010 {\it PNAS} \textbf{107}, 19844 

\bibitem{andabloreyes2005} Andablo-Reyes E, Díaz-Leyva P and Arauz-Lara J L
2005 {\it Phys. Rev. Lett.} \textbf{94}, 106001 

\bibitem{gutierrezsosa2018} Gutierrez-Sosa C, Merino-Gonzalez A, Sanchez R, Kozina A and Diaz-Leyva P 2018 {\it Macromolecules} \textbf{51}, 9203 

\bibitem{oliveira2019} Oliveira F A, Ferreira R M S, Lapas L C and Vainstein M H, 2019 {\it Front. Phys.} \textbf{7}, 1 

\bibitem{qian2003} Qian H 2003 {\it Processes with Long-Range Correlations: Theory and Applications} ed G Rangarajan and M Z Ding (Springer-Verlag), p 22.


\bibitem{figueroa2018} Figueroa-Morales N, Darnige T,  Douarche C, Martinez V, Soto R, Lindner A, and Clément E, E 2020 {\it Phys. Rev. X} \textbf{10} 021004

\bibitem{kou2004} Kou S C and Sunney Xie X 2004 {\it Phys. Rev. Lett.} \textbf{93}, 180603

\bibitem{dietrich1997}  Dietrich C R and Newsam G N 1997 {\it SIAM J. Sci. Comput.} \textbf{18}, 1088 

\bibitem{loewen2016} L\"owen H 2016 {\it Eur. Phys. J. Special Topics} \textbf{225}, 2319 

\bibitem{hu2019} Hu W-F, Lin  T-S , Rafai S and Misbah C 2019 {\it Phys. Rev. Lett.} \textbf{123}, 238004 

\bibitem{furutsu1963} Furutsu K 1963 {\it J. Res. Natl. Inst. Stand. Technol.} \textbf{67(D)}, 303 

\bibitem{novikov1965} Novikov E A 1965 {\it Sov. Phys. JETP} \textbf{20}, 1290 

\bibitem{haenggi1982} H\"anggi P and Thomas H 1982 {\it Phys. Rep.} \textbf{88}, 207

\bibitem{abramowitzBook} Abramowitz M and Stegun I A 1964 {\it Handbook of Mathematical Functions with Formulas, Graphs, and Mathematical Tables} ( New York: Dover Publications Inc.)

\bibitem{guinand1941} Guinand A P 1941 {\it Ann. Math.} \textbf{42}, (3), 591

\bibitem{applebaum2016} Applebaum D 2016 {\it Forum Mathematicum} \textbf{29}, (3), 501

\bibitem{sandev2014} Sandev T, Metzler R and Tomovski Z 2014 {\it J. Math. Phys.} \textbf{55}, 023301 

\bibitem{rodriguez2015}  Rodriguez R F, Fujioka J and Salinas-Rodriguez E 2015 {\it Physica A} \textbf{427}, 326 

\end{thebibliography}
\end{document}